\documentclass [pre,twocolumn,amsmath,amssymb,longbibliography]{revtex4-2}
\setcitestyle{numbers,square}
\usepackage{appendix}
\usepackage{array}
\usepackage{amsmath,amssymb}
\usepackage{tabularx}
\usepackage{graphicx}
\usepackage{bmpsize}
\usepackage{bm}
\usepackage{color}
\usepackage{amssymb}
\usepackage [version=3]{mhchem}
\usepackage{newtxmath}
\usepackage{hyperref}
\usepackage{url}
\usepackage{physics}
\usepackage{here}
\usepackage{float}
\usepackage{comment}

\hypersetup{
   unicode=true,          
   plainpages=false,
   colorlinks=true,       
   citecolor=blue,        
}

\DeclareMathAlphabet{\mathpzc}{OT1}{pzc}{m}{it}

\def\sgn{\mathrm {sgn}}

\newcommand{\rev} [1]{\textcolor{black}{ #1}}

\begin{document}

\title{Directional information transfer between interacting Brownian particles}



\author{Tenta Tani}
\affiliation{Department of Physics, The University of Osaka, Toyonaka, Osaka 560-0043, Japan}

\date{\today}

\begin{abstract}
We theoretically investigate how information flows when two particles interact with each other.
Understanding the physical mechanisms of directional information flow is crucial for advancing information thermodynamics and stochastic computing. 
However, the fundamental connection between mechanical motion and causal information transfer remains elusive. 
To focus only on essential effects of physical dynamics, we examine two interacting Brownian particles confined in a one-dimensional potential. 
By simulating their Langevin dynamics, we quantify the causal information exchange using transfer entropy. 
We demonstrate that a mass asymmetry inherently breaks the symmetry of information flow, inducing a net directional transfer from the heavier to the lighter particle.
Physically, the heavier particle, possessing larger inertia and higher active information storage, retains the memory of its trajectory longer against thermal fluctuations, thereby acting as a source of information.
We analytically clarify that this net transfer is governed by a competition between the difference in memory capacity and the predictability of the particle trajectories.
Furthermore, we reveal that the net information flow scales logarithmically with the mass ratio. 
These findings provide essential insights into the physical significance of transfer entropy and the nature of information flow in general physical systems.
\end{abstract}

\maketitle
\section{Introduction}
\label{sec:intro}
Directional flow of information has gained increasing importance across various scientific disciplines in recent years. 
In the context of information theory, the quantitative evaluation of directional information transfer was pioneered by Schreiber~\cite{Schreiber2000} through the introduction of transfer entropy (TE).
In Ref.~\cite{Schreiber2000}, the information flow between heartbeat and breathing was examined by the notion of TE; 
by comparing the magnitude of TE from the heart rate to the breathing rate with that in the reverse direction, a directional information flow was revealed.
Since then, this framework has been widely applied to analyze the dynamics of diverse complex systems, ranging from neural systems~\cite{Kobayashi2013,Kawasaki2014,Staniek2008,Vicente2011,Sabesan2009,Vakorin2011} to stock markets~\cite{Kwon2008,marschinski2002}.

In physics, information thermodynamics~\cite{parrondo2015thermodynamics,andrieux2008nonequilibrium,jarzynski2008thermodynamics,hayashi2010fluctuation,sagawa2008second,sagawa2009minimal,sagawa2010generalized,sagawa2012fluctuation,Sagawa2013role,karbowski2024information,horowitz2014} has become increasingly important in recent years.
One of the most intriguing consequences of information thermodynamics is the equivalence of information and work, as demonstrated by the resolution of Maxwell’s demon paradox~\cite{Szilard1929,brillouin1951maxwell,Bennett1982,landauer1961irreversibility,sagawa2008second,sagawa2009minimal,sagawa2010generalized,sagawa2012fluctuation,Sagawa2013role,Koski2014,Goerlich2025,Lagoin2022}.
An information-theoretical quantity called mutual information is associated with work; the TE is a type of mutual information.
The TE, a measure of information flow, is also an important quantity in information thermodynamics~\cite{Rozo2021,sandoval2014structure,falkowski2023causality,sun2014identification,gao2020single,murcio2015urban,ito2016backward,oka2013exploring,ito2013information,ito2015informationflowentropyproduction,Ito2015,Spinney2016,Potts2018,Horowitz2014secondlawlike,Auconi2019,Horowitz2015,Ruizpino2026}.

From a practical standpoint, information flow is crucial in the emerging field of stochastic computing~\cite{Bennett1982,pinna2018skyrmion,zazvorka2019thermal,ishikawa2021implementation}.
Despite the growing attention to such applications, the fundamental relationship between the underlying physical motion of particles and the resulting information flow has not yet been elucidated.
It is not clear how information is transferred through particle collisions, and more generally, how we can interpret the TE in physical systems.
By elucidating these problems in simpler physical systems, we can apply information flow concepts to various physical systems including nanoparticles in liquids and molecules in gases.
Ultimately, studies of the TE in these systems are essential to strengthen the connection between information theory and physics.

\rev{Previously, the stochastic dynamics of interacting particle systems have been investigated.
For instance, studies have revealed how noise correlations and harmonic couplings affect the escape dynamics of Brownian dimers~\cite{Singh2017}, and how inter-particle interactions modulate the many-body Arrhenius law~\cite{Kumar2024_PRE, Kumar2024_JCP, Kumar2025_JCP}.
While these studies have deepened our thermodynamic understanding of escape rates and macroscopic fluctuations driven by interactions, the fundamental relation between such mechanical interactions and the causal flow of information remains less explored.}

Recently, we examined the meaning and role of the TE in magnetic skyrmions confined in a box both theoretically~\cite{tani2026} and experimentally~\cite{suzuki2025informationdynamicsnaturalcomputing}, motivated by their potential application as information carriers, or ``skyrmion bits"~\cite{ishikawa2021implementation,suzuki2025informationdynamicsnaturalcomputing} towards ultralow-power stochastic computing.
However, in these studies, the system parameters were fixed in principle. 
Most notably, the two interacting particles were treated as identical, which naturally leads to a perfectly symmetric information exchange, i.e., $T_{X\to Y} = T_{Y\to X}$.
Furthermore, skyrmion dynamics inevitably involve gyrotropic motion originating from their nontrivial topological properties~\cite{miki2021brownian,zhao2020topology,suzuki2021diffusion}. 
This inherent complexity obscures the underlying fundamental connection between mechanical motion and information transfer. 
While efficiency of information flow between nonequilibrium Brownian particles driven by a temperature difference has been studied~\cite{Allahverdyan2009}, the purely mechanical effects remain unexplored. 
Specifically, it is still unknown how a difference in inertia (mass) between interacting particles under a uniform thermal environment dictates the causal directionality of information transfer.

In this paper, to reveal the essential relationship between physical motions and information transfer, we focus on one of the simplest setups: two interacting asymmetric Brownian particles confined within a one-dimensional wall potential.
By describing the system with coupled Langevin equations and quantifying the causal information flow via TE, we demonstrate that a mass asymmetry inherently induces a directional asymmetry in the information exchange.
We reveal that net information (the difference of TEs in both directions, $T_{Y\to X} - T_{X\to Y}$) predominantly flows from the particle with larger inertia to the one with smaller inertia.
The heavier particle is less susceptible to thermal fluctuations and thus retains the memory of its past trajectory for a longer duration. 
It acts as a stable source of information, which is quantitatively supported by its higher active information storage (AIS)~\cite{LIZIER201239}.
Furthermore, we analytically establish a direct mathematical connection between the directional information flow and the memory capacity.
By decomposing the net information flow, we explicitly show that this causal directionality is governed by a competition between the difference in AIS and the difference in the predictability of the particle trajectories (conditional Shannon entropy).
Crucially, we clarify that conventional symmetric measures, such as mutual information and correlation functions, merely capture shared mechanical periodicity and fail to detect this causal directionality.

Moreover, we systematically evaluate the mass-ratio dependence of the net information flow, finding that the temporal delay of the information transfer increases with the inertia. 
Importantly, we establish that the net information flow scales logarithmically with the mass ratio.
These results offer a fundamental understanding of the physical meaning of TE and directed information transfer in general physical systems.

The remainder of this paper is organized as follows.
In Sec.~\ref{sec:formulation}, a model of interacting two Brownian particles is introduced.
We nondimensionalize the equations of motion to simplify the system description.
Section \ref{sec:result} presents the simulated system dynamics and analyses using information theory.
We reveal the nature of net information propagation, which is induced by mass asymmetry.
Finally, Sec.~\ref{sec:conclusion} concludes the paper.

\section{Model of Brownian particles in one-dimensional box}
\label{sec:formulation}

\begin{figure}
\begin{center}
   \includegraphics [width=0.9\linewidth]{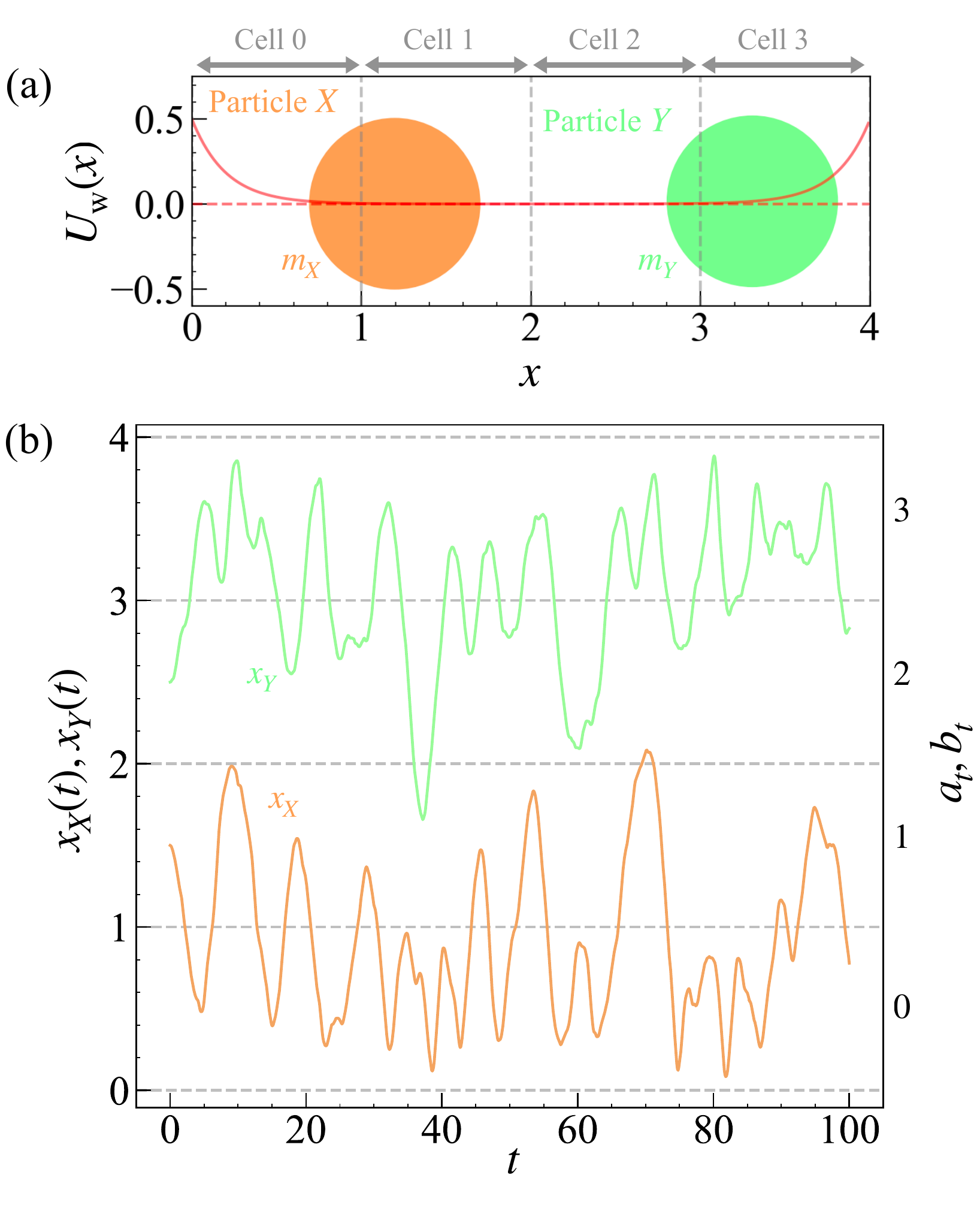}
   \caption{
   (a)~Snapshot of two interacting Brownian particles ($X$ and $Y$) in a wall potential.
   The curve illustrates the potential from the two walls, which vanishes in $1 \leq x/\xi \leq 3$.
   (b)~Trajectories of the two particles, which are well confined between the walls.
   \rev{The right vertical axis illustrates how the continuous positions $x_X(t)$ and $x_Y(t)$ are mapped into the discrete states $a_t$ and $b_t$.}
   \rev{Because of the repulsive interaction, particle $X$ ($Y$) predominantly exists in the left (right) region. 
   However, it should be noted that due to the presence of thermal noise and the softness of the interaction potential, the two particles can occasionally cross each other during rare fluctuation events.}
            }
\label{fig:dynamics}
\end{center}
\end{figure}

\subsection{Equations of motion}
We consider two interacting Brownian particles, \rev{$X$ and $Y$.}
Their one-dimensional motions are governed by Langevin equations,
\rev{
\begin{equation}
    \begin{split}
        m_X \dv[2]{x_X}{t} &= -\gamma \dv{x_X}{t} + F_\mathrm{int}(x_X,x_Y) + F_\mathrm{w}(x_X) + R_X,
        \\
        m_Y \dv[2]{x_Y}{t} &= -\gamma \dv{x_Y}{t} - F_\mathrm{int}(x_X,x_Y) + F_\mathrm{w}(x_Y) + R_Y,
    \end{split}
    \label{eq:langevin}
\end{equation}
}
\rev{where $x_X$ ($x_Y$) is the position of the particle $X$ ($Y$).}
The parameter $\gamma$ is a damping coefficient, assumed to be identical for both particles.
We model the particle--particle interaction using an exponentially decaying repulsive potential,
\rev{
\begin{equation}
    U_\mathrm{int}(x_X, x_Y) = U_0 e^{-|x_X-x_Y|/\xi},
\end{equation}
}
where $U_0$ is the interaction energy and $\xi$ is the interaction range.
\rev{The force $F_\mathrm{int}(x_X, x_Y)$ represents the interaction force exerted on particle $X$ by particle $Y$, which is derived from the potential $U_\mathrm{int}(x_X, x_Y)$ as follows:}
\rev{
\begin{equation}
\begin{split}
    F_\mathrm{int}(x_X,x_Y) &= -\pdv{U_\mathrm{int}}{x_X} \\ &= \frac{U_0}{\xi}e^{-|x_X-x_Y|/\xi} \sgn{(x_X-x_Y)}.
\end{split}
\end{equation}
}
Here, $\sgn(x)$ denotes the sign function.
We assume the walls at $x=0$ and $L$ that yield an exponentially decaying potentials as well,
\begin{equation}
    U_\mathrm{w}(r) = U_{\mathrm{w}0} [ e^{-r/\xi_\mathrm{w}} + e^{-(L-r)/\xi_\mathrm{w}} ],
\end{equation}
where $U_{\mathrm{w}0}$ and $\xi_\mathrm{w}$ are the energy scale and interaction range of the wall potential, respectively.
The symbol $r$ denotes the particle coordinate.
The corresponding wall force is explicitly written as
\begin{equation}
    F_\mathrm{w}(r) = \frac{U_{\mathrm{w}0}}{\xi_\mathrm{w}} [ e^{-r/\xi_\mathrm{w}} - e^{-(L-r)/\xi_\mathrm{w}} ].
\end{equation}
The last terms in Eq.~\eqref{eq:langevin}, \rev{$R_X$} and \rev{$R_Y$}, represent the random force originating from thermal fluctuations.
\rev{Note that $R_X$ and $R_Y$ represent independent Gaussian white noise terms satisfying $\langle R_j(t) \rangle = 0$ and $\langle R_j(t) R_k(t') \rangle = 2\gamma k_\mathrm{B}T \delta_{jk} \delta(t-t')$, where $j, k \in \{X, Y\}$.
They are discretized in our numerical simulations as
\begin{equation}
    R_X,R_Y \sim \sqrt{\frac{2\gamma k_\mathrm{B}T}{\delta t}} \mathcal{N}(0,1),
\end{equation}
where $\mathcal{N}(0,1)$ is the standard normal distribution.
Here, $\delta t$ represents the discrete time step.
} 

Here, we model both the particle--particle and particle--wall interactions using exponentially decaying functions for both physical and computational reasons. 
Physically, these exponential forms generically capture soft short-range repulsions, which are realized in interacting quasiparticles such as magnetic skyrmions~\cite{tamura2020skyrmion,miki2021size,tani2026,capic2020skyrmion,lin2013particle}, for instance. 
Computationally, unlike hard-core or algebraically diverging potentials (e.g., the Lennard--Jones potential), exponential potentials do not exhibit singularities at zero distance. 
This softness is highly advantageous in Langevin dynamics simulations, as it prevents unphysical numerical instabilities that could otherwise occur when random thermal fluctuations occasionally drive the particles extremely close to each other or to the walls.

To simplify the form and concentrate on the fundamental physics, we nondimensionalize the Langevin equations [Eq.~\eqref{eq:langevin}].
We choose \rev{$m_X$}, $U_0$, and $\xi$ as the basic units of mass, energy, and length, respectively.
To introduce mass asymmetry, we set \rev{$m_X\neq m_Y$}, which yields a mass ratio \rev{$\mu \equiv m_Y/m_X \neq 1$}.
A detailed description of this procedure is given in Appendix~\ref{app:derivation}.

\subsection{Particle dynamics}
Figure~\ref{fig:dynamics}(a) shows a schematic picture of two Brownian particles (particle $X$ and $Y$) confined in a box. 
The vertical axis is the wall potential $U_\mathrm{w}$ with \rev{$L=4$}.
In order to investigate the effect of interaction, we use parameter values \rev{$U_{\mathrm{w}0} = 0.5$} and \rev{$\xi_\mathrm{w} = 0.2$}, which renders the wall potential negligibly small ($U_\mathrm{w} \approx 0$) in the central range of \rev{$1 \leq x \leq 3$}.

\rev{In the numerical simulation of Eq.~\eqref{eq:langevin}, we use the fourth-order Runge--Kutta method, which corresponds to the Stratonovich calculus~\cite{Rumelin1982}.}
Since the coefficient of the random force is constant (the system is driven by additive noise), the It\^{o} and Stratonovich calculus yield the same results.
The simulation time is \rev{$t_\mathrm{f} = 100$} with time increment of \rev{$\delta t = 0.1$}.
We employ the temperature parameter \rev{$T = 0.1$}, and the damping constant is fixed to be \rev{$\gamma = 0.3$}.
In Fig.~\ref{fig:dynamics}(b), we illustrate the dynamics of the system with $\mu=1$ (two identical particles).
At $t=0$, the particles are released from rest at \rev{$x_X = 1.5$} and \rev{$x_Y = 2.5$}.
As shown in Fig.~\ref{fig:dynamics}(b), the two particles are well confined within the box (\rev{$0\leq x \leq 4$}) and clearly exhibit mutual repulsion.

Throughout the paper, we fix the above parameters \rev{$(\gamma, U_{\mathrm{w}0}, \xi_\mathrm{w}, L, T)$} while varying the mass ratio $\mu$, which serves as the sole source of symmetry breaking between the particles.
\rev{Note that, in actual experiments, typically using colloidal or nanoparticles suspended in a fluid~\cite{Crocker1994,Toyabe2010,Berut2012}, both the particle mass and the viscous friction coefficient are determined by the particle size and therefore cannot generally be tuned independently~\cite{Franosch2011,Huang2011}.}
However, if both parameters were varied simultaneously, it would be difficult to disentangle whether the asymmetric information flow originates from the difference in inertia or the difference in damping coefficient. 
Therefore, to unambiguously isolate the purely mechanical effect of inertia, we theoretically treat the friction and the mass as independent parameters and assign an identical \rev{$\gamma$} to both particles. 
This idealization allows us to establish that mass asymmetry alone is fundamentally sufficient to induce a directed causal information transfer.

\section{Information-theoretical analysis}
\label{sec:result}
\subsection{Shannon entropy}
\label{subsec:shannon}

To analyze the flow of information between the particles, we numerically simulate the Langevin equations [Eq.~\eqref{eq:langevin-dimless}] $N=10^6$ times.
This ensemble size sufficiently suppresses statistical errors and enables the following analyses of information flow.
From each \rev{trajectory}, we obtain two time series,
\rev{
\begin{equation}
\begin{split}
    &\{x_X^{(i)}(0), x_X^{(i)}(\delta t), x_X^{(i)}(2 \delta t), \cdots, x_X^{(i)}(t_\mathrm{f})\},
    \\ 
    &\{x_Y^{(i)}(0), x_Y^{(i)}(\delta t), x_Y^{(i)}(2 \delta t), \cdots, x_Y^{(i)}(t_\mathrm{f})\},
\end{split}
\end{equation}
}
where $i=1,2,\cdots,N$ specifies an individual \rev{trajectory}.

For the calculation of entropic quantities, we discretize the position as
\rev{
\begin{equation}
    \begin{split}
        &\text{Cell 0: } x<1
        \\
        &\text{Cell 1: } 1 \leq x < 2
        \\
        &\text{Cell 2: } 2 \leq x < 3
        \\
        &\text{Cell 3: } x > 3
    \end{split}
\end{equation}
}
as shown in Fig.~\ref{fig:dynamics}(a).
\rev{Here, the spatial discretization is introduced to robustly evaluate the probability distributions required for the information-theoretic quantities. 
A detailed discussion justifying the choice of four cells and the physical properties of the continuous limit is provided in Sec.~\ref{subsec:TE}.}
Because of the wall potential, the particles rarely move inside the walls (\rev{$x < 0$} and \rev{$x > 4$}).
Under this criterion, after executing $N$ simulations, we discretize the positions as
\rev{
\begin{equation}
    \begin{split}
        &x_X^{(i)}(t) \to a_t^{(i)} \in \{0,1,2,3\},
        \\
        &x_Y^{(i)}(t) \to b_t^{(i)} \in \{0,1,2,3\},
    \end{split}
\end{equation}
}
where $a_t$ and $b_t$ are the discrete spatial states of particle $X$ and $Y$, respectively. 
We define the corresponding stochastic variables as $X_t$ and $Y_t$, which take the values $a_t$ and $b_t$.

The probability that particle $X$ exists in cell $a_t$ at time $t$ is given by
\begin{equation}
    p(a_t) = \frac{1}{N}\sum_{i=1}^{N} \rev{\delta_{a_{t}^{(i)}, a_{t}}},
\end{equation}
where \rev{$\delta_{i,j}$} is Kronecker's delta,
\rev{
\begin{equation}
\delta_{i,j} =
\begin{cases}
1 & (i=j) \\
0 & (i \neq j).
\end{cases}
\end{equation}
}
Similarly, the joint probability between different times $t$ and $t'$ is written as
\rev{
\begin{equation}
    p(a_t, a_{t'}) = \frac{1}{N}\sum_{i=1}^{N} \delta_{a_{t}^{(i)}, a_{t}} \delta_{a_{t'}^{(i)}, a_{t'}},
\end{equation}
}
and the conditional probability is obtained from the relation $p(a_t \mid a_{t'}) = p(a_t, a_{t'}) / p(a_{t'})$.

\begin{figure}
\begin{center}
   \includegraphics [width=0.8\linewidth]{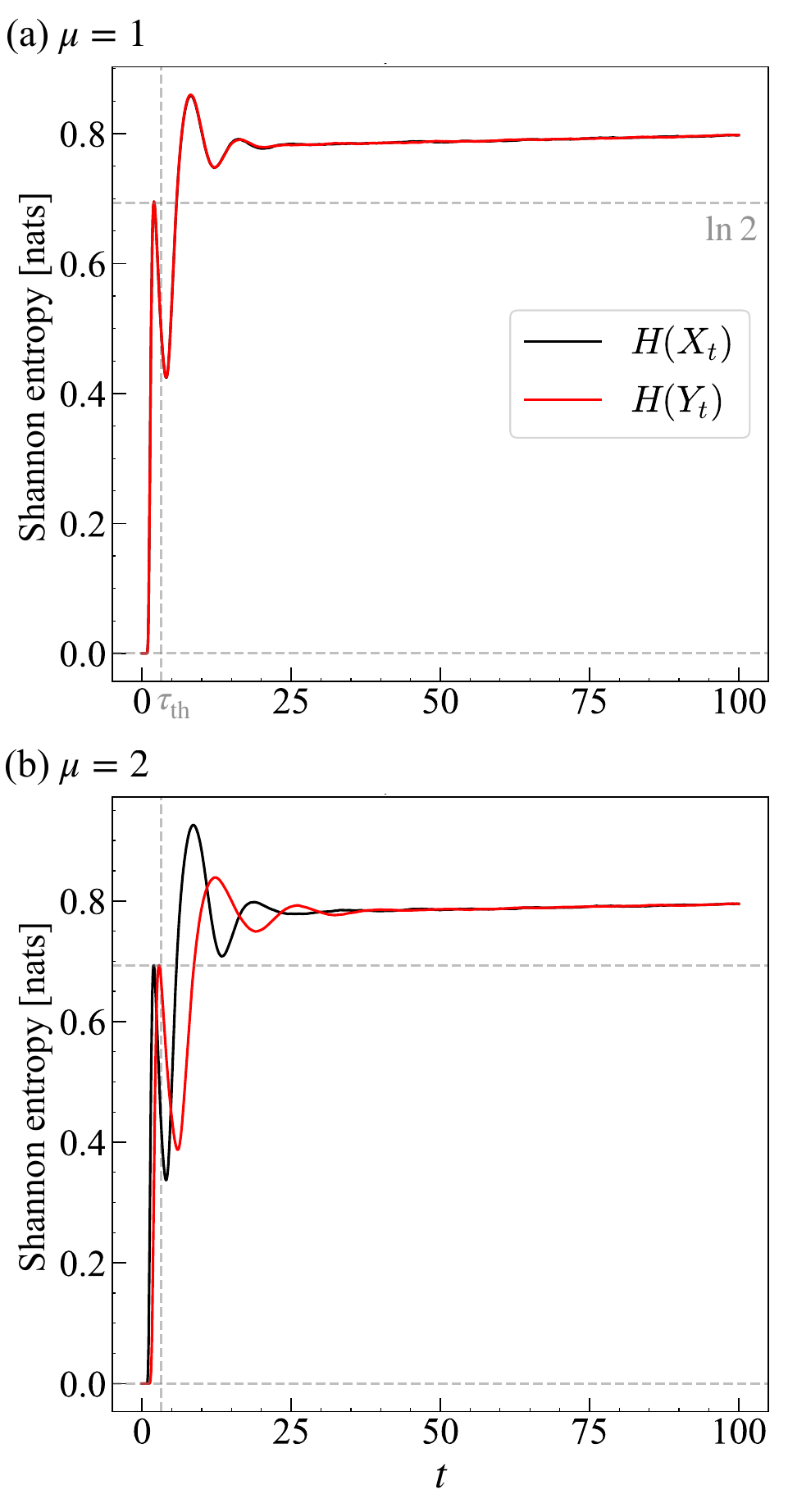}
   \caption{
   (a)~Time evolution of Shannon entropy $H(X_t)$ (particle $X$) and $H(Y_t)$ (particle $Y$) for mass parameter $\mu = 1$ (identical particles).
   {The vertical line, $\tau_\mathrm{th} = \xi / v_\mathrm{th}$, represents the characteristic time of thermal motion (see the text).}
   (b)~Similar plots for an asymmetric case $\mu=2$ ($Y$ is twice heavier than $X$).
            }
\label{fig:shannon}
\end{center}
\end{figure}

For particles $X$ and $Y$, the Shannon entropy at time $t$ is written as
\begin{equation}
\begin{split}
    H(X_t) &= -\sum_{a_t=0}^3 p(a_t) \ln{p(a_t)},
    \\
    H(Y_t) &= -\sum_{b_t=0}^3 p(b_t) \ln{p(b_t)},
\end{split}
\end{equation}
which represents the randomness of a particle's position.
Figure~\ref{fig:shannon} illustrates the time evolution of Shannon entropies $H(X_t), H(Y_t)$ for $\mu = 1$ and $2$.
In Fig.~\ref{fig:shannon}(a), we show the symmetric case, $\mu=1$. 
As expected, the two entropy curves are almost perfectly identical.
The initial rapid increase in $H$ corresponds to the expansion of the probability distribution from a deterministic initial state that is driven by the repulsive interaction and thermal fluctuations.
The subsequent oscillations reflect the following processes: the expanding distribution hits the soft wall potentials and bounces back, causing a transient spatial localization of the particles.
The vertical dashed line indicates the characteristic time scale $\tau_\mathrm{th} \equiv \xi/v_\mathrm{th}$~\cite{tani2026}, where $v_\mathrm{th}$ is the thermal velocity defined as
\rev{
\begin{equation}
    v_\mathrm{th} = \sqrt{\frac{k_\mathrm{B}T}{m_X}}.
\end{equation}
}
It can be seen that this rebound process occurs on the order of $\tau_\mathrm{th}$.
After the initial transient oscillations are damped out (\rev{$t \gtrsim 25$}), the Shannon entropies reach a steady state, which gives $H \approx 0.8$.
This value is slightly larger than $\ln{2}$ (horizontal dashed line); although a particle is predominantly localized in two out of the four cells (i.e., cell 0 and 1 for particle $X$), it sometimes has a chance to explore the adjacent cell (cell 2).
The effective volume occupied by a single particle can be estimated as \rev{$L_\mathrm{eff} = e^{0.8} \approx 2.2$} cells.

Figure~\ref{fig:shannon}(b) shows a similar plot for the asymmetric case, $\mu=2$.
We clearly see that the oscillation of particle $Y$ (heavier particle) becomes slower with a smaller amplitude compared to particle $X$.
This represents its larger inertia.
The probability distribution of the heavier particle expands and localizes more gradually than that of the lighter one.
The relaxation process takes longer than in the symmetric case; both particles reach a steady state for \rev{$t \gtrsim 40$}. 
\rev{This delayed relaxation originates from the larger absolute mass of particle $Y$, which slows down its own dynamics and, through the particle--particle interaction, consequently delays the overall relaxation of the coupled system.}

\subsection{Mutual information and active information storage}
\label{subsec:mi-ais}

\rev{To understand the fundamental relationship between the interacting particles, it is first necessary to quantify the degree of correlation between their states.
Unlike conventional correlation functions $C(\tau)$ that capture only linear dependences (see Appendix~\ref{app:corr}), mutual information quantifies the total amount of shared information, including nonlinear dependencies.
Therefore, by calculating the mutual information $I(X_t : Y_{t-\tau})$, we examine how the states of the two particles depend on each other over a time delay $\tau$.}
The mutual information is defined as
\rev{
\begin{equation}
\label{eq:MI-definition}
\begin{split}
    I(X_t : Y_{t-\tau}) &= H(X_t) - H(X_t \mid Y_{t-\tau})
    \\ &= 
    \sum_{a_t, b_{t-\tau}=0}^3 p(a_t, b_{t-\tau}) \ln{\frac{p(a_t, b_{t-\tau})}{p(a_t)p(b_{t-\tau})}},
\end{split}
\end{equation}
}
where conditional Shannon entropy \rev{$H(X_t \mid Y_{t-\tau})$} is given by
\rev{\begin{equation}
    H(X_t \mid Y_{t-\tau}) = -\sum_{a_t, b_{t-\tau}=0}^3 p(a_t, b_{t-\tau}) \ln{p(a_t \mid b_{t-\tau})}.
\end{equation}}
Mutual information \rev{$I(X_t : Y_{t-\tau})$} represents information shared between particle $X$ at time $t$ ($X_t$) and particle $Y$ at time \rev{$t-\tau$ ($Y_{t-\tau}$).}
For large time delay \rev{$\tau$}, $X_t$ and \rev{$Y_{t-\tau}$} become uncorrelated, giving \rev{$H(X_t \mid Y_{t-\tau}) = H(X_t)$} and therefore \rev{$I(X_t : Y_{t-\tau}) = 0$} (no shared information).
\rev{In other words, the motions of particle $X$ and $Y$ become independent, i.e., $p(a_t, b_{t-\tau}) = p(a_t) p(b_{t-\tau})$. 
Therefore, we obtain $I(X_t : Y_{t-\tau})=0$ from Eq.~\eqref{eq:MI-definition}.
}

\begin{figure}
\begin{center}
   \includegraphics [width=0.9\linewidth]{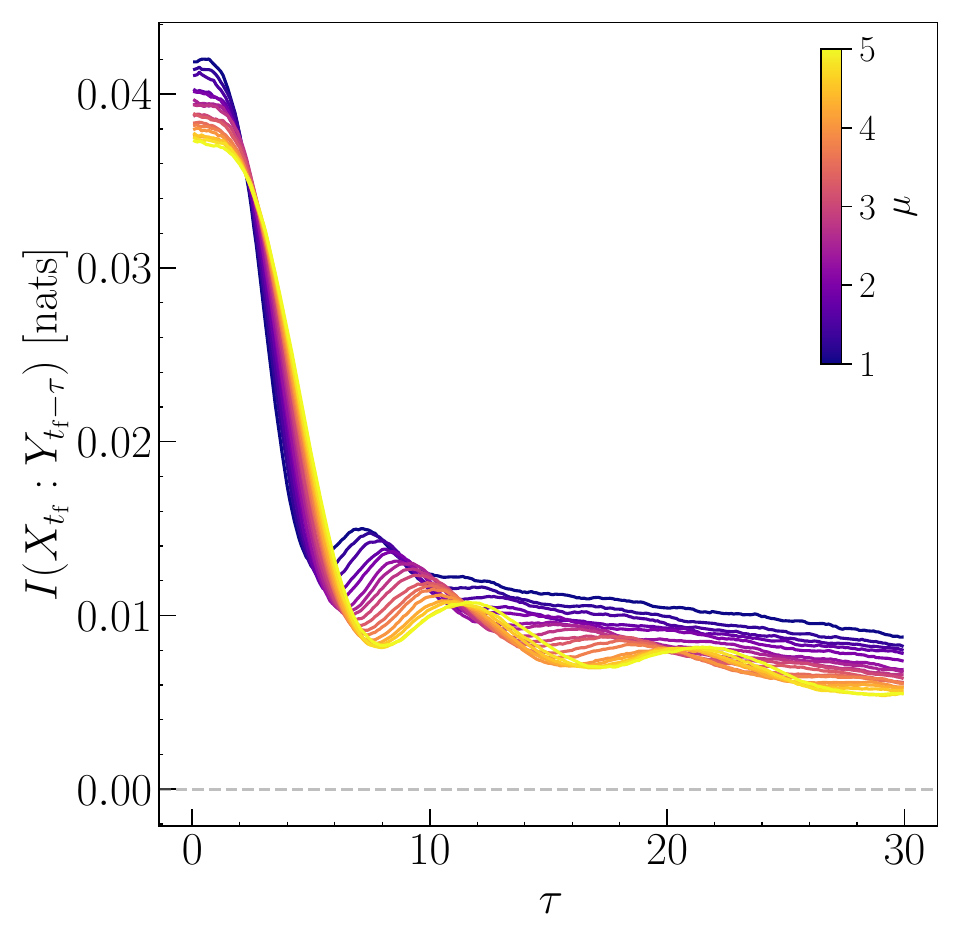}
   \caption{
   Mutual information $I(X_t : Y_{t-\tau})$ at the reference time $t=t_\mathrm{f}$, illustrated as a function of the time lag $\tau$.
   The mass ratio $\mu$ varies from $\mu=1$ to $5$.
            }
\label{fig:MI}
\end{center}
\end{figure}
Figure~\ref{fig:MI} illustrates the \rev{$\tau$} dependence of the mutual information for several values of mass ratio $\mu$.
In order to examine steady states, we fix the reference time to $t = t_\mathrm{f}$.
\rev{Because this reference time $t_\mathrm{f} = 100$ is sufficiently larger than the relaxation time of the system [e.g., $t \approx 40$ for $\mu=2$, as seen in Fig.~\ref{fig:shannon}(b)], the system has completely lost the memory of its initial state. Therefore, the steady-state results presented hereafter, including the magnitude and directionality of the information flow, are independent of the specific choice of initial particle positions.}
The stochastic motion of either particle degrades the shared information over time.
Thus, the mutual information, rapidly decays as the time lag \rev{$\tau$} increases.
Crucially, the decay rate strongly depends on the mass ratio. 
For the symmetric case ($\mu=1$), the mutual information exhibits a relatively fast decay accompanied by a transient oscillation around \rev{$\tau \approx 7.5$}, reflecting the rapid rebound process against the walls (see Sec.~\ref{subsec:shannon}). 
\rev{In contrast, as $\mu$ increases, the heavier particle $Y$ possesses a larger absolute mass, which significantly slows down its dynamics.}
Consequently, the memory of its past state is preserved for a longer duration, leading to a much slower decay of the mutual information.

As pointed out in Ref.~\cite{horowitz2014}, \rev{$[I(X_t : Y_{t-\tau}) - I(X_t : Y_t)] / \tau$} expresses an internal information flow, which realize an observation and a feedback process in the system (Maxwell’s demon’s task). 
In our system, the initial slopes in Fig.~\ref{fig:MI} is zero, showing the absence of the Maxwell’s demon.

While these observations provide physical insights into the correlation times of the system, the mutual information \rev{$I(X_{t_\mathrm{f}} : Y_{t_\mathrm{f} - \tau})$} itself is fundamentally a symmetric measure; it only quantifies the shared information between two variables and cannot distinguish whether the correlation arises from $X$ influencing $Y$, or $Y$ influencing $X$. 
Therefore, the mutual information lacks the ability to detect the directionality of information flow~\cite{Schreiber2000}.
Similarly, a conventional correlation function only captures the shared mechanical periodicity and fails to reveal the causal directionality (see Appendix~\ref{app:corr} for detailed discussions).
To reveal the true causal influence and the directional flow of information driven by the mass asymmetry, we must turn to a more sophisticated measure, TE.

\begin{figure}
\begin{center}
   \includegraphics [width=0.85\linewidth]{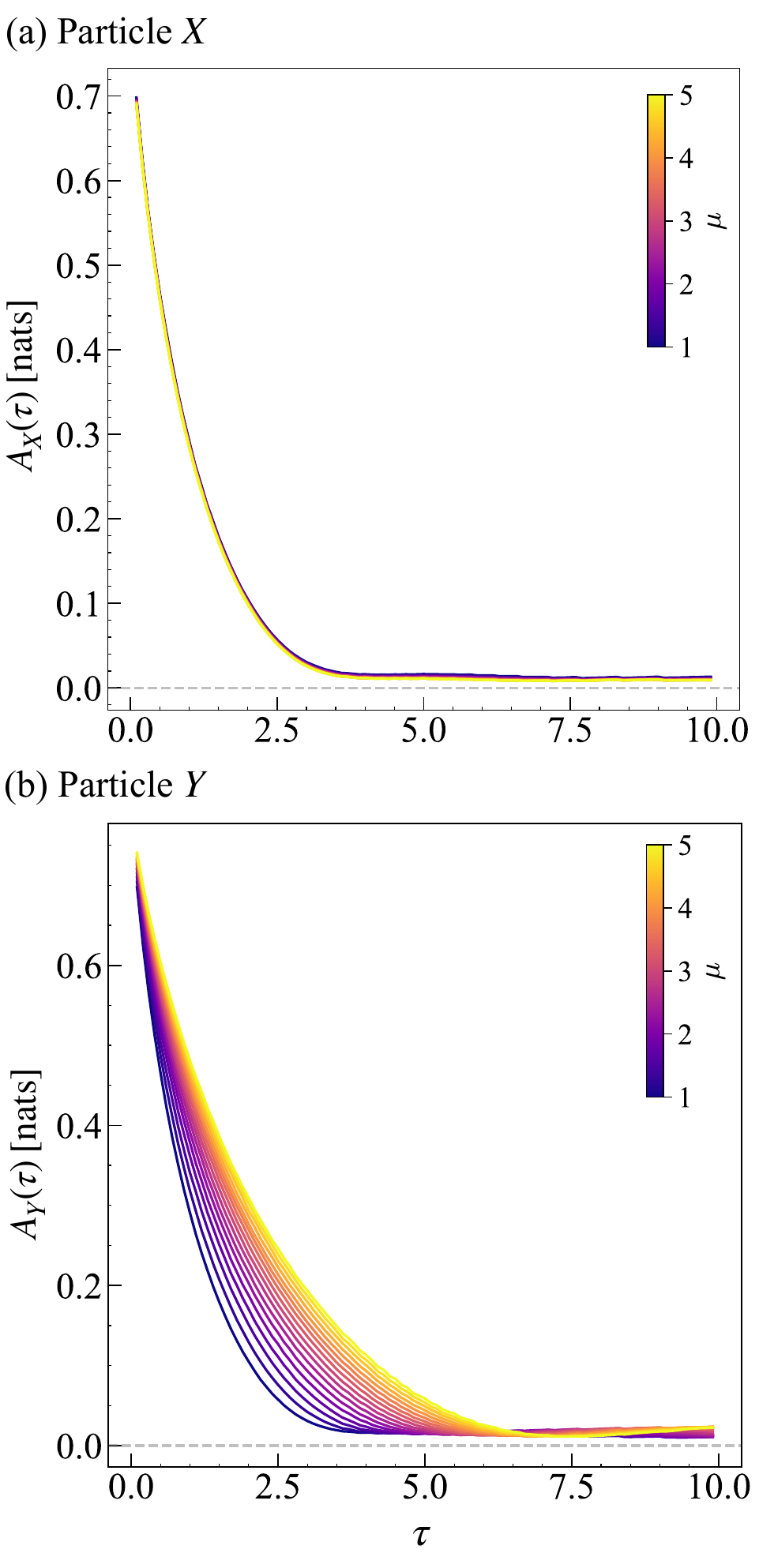}
   \caption{
   Active information storage of particle $X$ [panel (a), lighter] and $Y$ [panel (b), heavier], as a function of the time lag $\tau$.
   The mass ratio $\mu$ is varied from 1 to 5.
            }
\label{fig:AIS}
\end{center}
\end{figure}
Before analyzing the information transfer between the two particles using the TE, it is instructive to first quantify how much information each particle internally retains about its own past. 
For this purpose, we evaluate the AIS~\cite{LIZIER201239}, 
\rev{
\begin{equation}
\label{eq:ais-definition}
    \begin{split}
        A_X(\tau)&= I(X_t : X_{t-\tau}), \\
        A_Y(\tau)&= I(Y_t : Y_{t-\tau}),
    \end{split}
\end{equation}
}
which measures the predictable information from a particle's past state to its present state.
Figure~\ref{fig:AIS} shows the AIS for both particles as a function of the lag \rev{$\tau$} at the reference time $t=t_\mathrm{f}$.
{As the time lag \rev{$\tau$} increases, from the initial values \rev{$A_X(\tau=0) = H(X_t)$} and \rev{$A_Y(\tau=0) = H(Y_t)$}, the AIS drops to zero.}
As the mass ratio $\mu$ increases, the AIS of the lighter particle $X$ [Fig.~\ref{fig:AIS}(a)] remains almost unchanged, whereas the decay rate for the heavier particle $Y$ [Fig.~\ref{fig:AIS}(b)] decreases.
In other words, the heavier particle $Y$ consistently exhibits a higher AIS than the lighter particle $X$. 
This quantitatively confirms our previous observation from the mutual information: the larger inertia of $Y$ can retain a stronger memory of its own trajectory.

\subsection{Transfer entropy}
\label{subsec:TE}

\begin{figure*}
\begin{center}
   \includegraphics [width=\linewidth]{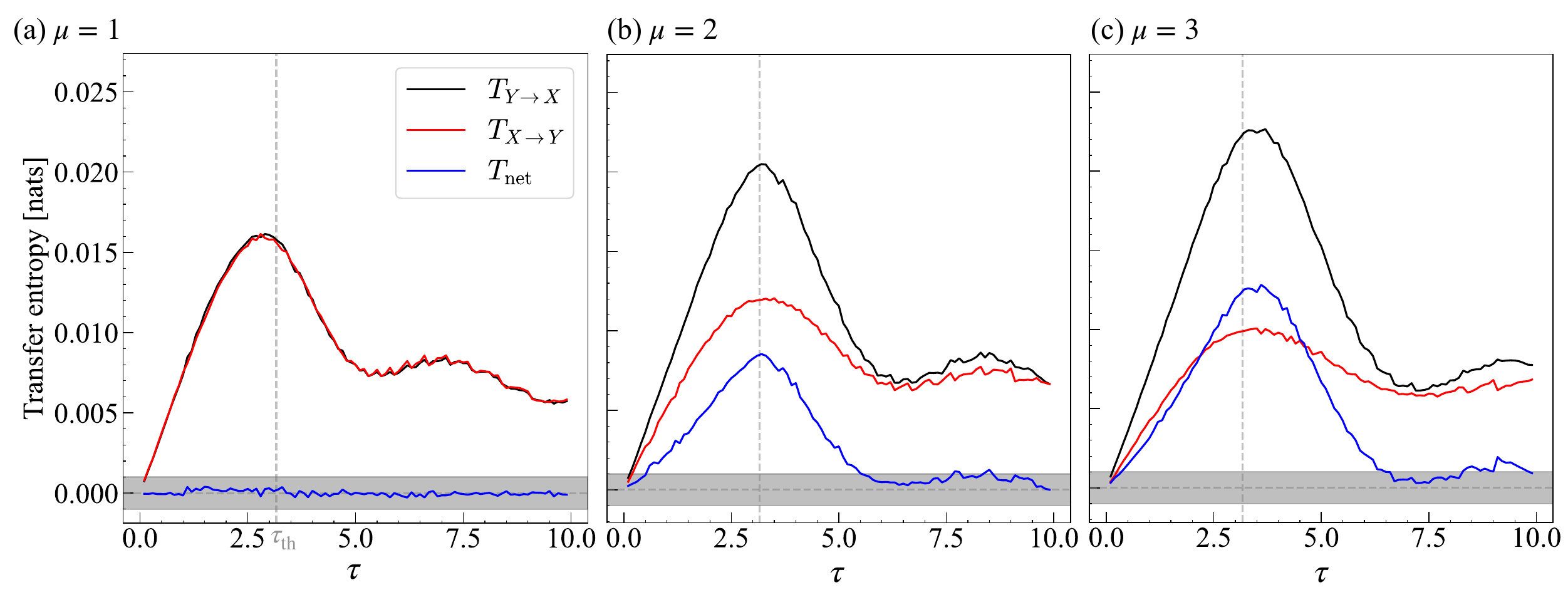}
   \caption{
   Transfer entropies $T_{Y\to X}$, $T_{X\to Y}$, and their difference $T_\mathrm{net}$ for (a)~$\mu=1$, (b)~$\mu=2$, and (c)~$\mu=3$. 
   The shaded region illustrates the statistical noise level $\pm 1/\sqrt{N}$.
   The vertical dashed lines represent the characteristic time scale of thermal motions, $\tau_\mathrm{th}$ (see the text).
            }
\label{fig:te-components}
\end{center}
\end{figure*}
Having established that the heavier particle (particle $Y$) acts as a more stable source of information (higher AIS) while the lighter particle (particle $X$) is more susceptible to thermal fluctuations, the crucial question arises: does this asymmetry in memory lead to a directional flow of information between them?
To answer this, we calculate the TE, which quantifies the amount of directed information transferred from a source variable to a target variable.
The TE from particle $Y$ to particle $X$ with a time lag \rev{$\tau$} is defined as
\rev{
\begin{align}
\begin{split}
    T_{Y \to X}(\tau) &= I(X_t : Y_{t-\tau} \mid X_{t-\tau})
    \\
    &= H(X_t \mid X_{t-\tau}) - H(X_t \mid X_{t-\tau}, Y_{t-\tau}),
\label{eq:TE}
\end{split}
\end{align}
}
which measures how much the past state of $Y$ (at time \rev{$t-\tau$}) reduces the uncertainty about the present state of $X$ (at time $t$), given the past state of $X$ itself.
\rev{
To explicitly clarify why this expression captures the directional information flow from $Y$ to $X$, it is instructive to examine each term. 
The first term $H(X_t \mid X_{t-\tau})$ represents the uncertainty about the present state of $X$ given only its own past history. 
The second term $H(X_t \mid X_{t-\tau}, Y_{t-\tau})$ represents the remaining uncertainty of $X$ when the past state of $Y$ is additionally known. 
Therefore, their difference quantifies the net information gained about $X_t$ that is provided by $Y_{t-\tau}$, beyond what is already predictable from $X$'s own past. 
This formulation fundamentally captures the causal influence directed from $Y$ to $X$.
Importantly, in general, this quantity is not symmetric between $X$ and $Y$, i.e., $T_{Y \to X} \neq T_{X \to Y}$.
}
The TE is a widely used measure of information flow~\cite{Kobayashi2013,Kawasaki2014,Staniek2008,Vicente2011,Rozo2021,Kwon2008,marschinski2002,sandoval2014structure,falkowski2023causality,sun2014identification,gao2020single,murcio2015urban,ito2016backward,oka2013exploring,ito2013information,Udoy2021,Sabesan2009,Vakorin2011,Brown2020,Basak2023}.
From the TE in both directions, \rev{$T_{Y \to X}(\tau)$} and \rev{$T_{X \to Y}(\tau)$}, we can calculate the net TE~\cite{Udoy2021,marschinski2002,Sabesan2009,Vakorin2011,Basak2023,Biswajit2025},
\rev{
\begin{equation}
\label{eq:def-net-te}
T_\mathrm{net}(\tau) = T_{Y \to X}(\tau) - T_{X \to Y}(\tau),
\end{equation}
}
to determine the dominant direction of the information flow.
{
While Eq.~\eqref{eq:def-net-te} provides the operational definition of the net information flow, its fundamental significance is governed by an information-theoretic sum rule, Eq.~\eqref{eq:shannon-decomposition}.
In a stationary state, the total uncertainty (Shannon entropy) of each particle is decomposed into its internal memory (AIS), the incoming information flow (TE), and the conditional Shannon entropy [Eq.~\eqref{eq:shannon-decomposition}].
As we will derive in Eq.~\eqref{eq:T-net-ais}, the mirror symmetry of our system enforces a balance between these components, revealing that $T_\mathrm{net}$ is intrinsically driven by the asymmetry in the particles' memory capacities.
}

Figure~\ref{fig:te-components} displays the time-lag dependence of the TEs, $T_{Y \to X}$ and $T_{X \to Y}$, along with their difference (net TE) $T_\mathrm{net}$, for three selected mass ratios: (a) $\mu=1$, (b) $\mu=2$, and (c) $\mu=3$.
Similar to the analysis in Sec.~\ref{subsec:mi-ais}, we fix the reference time to $t=t_\mathrm{f}$ to examine the steady-state information transfer.
For the identical particles [Fig.~\ref{fig:te-components}(a)], the two curves $T_{Y \to X}$ and $T_{X \to Y}$ almost overlap.
As a result, $T_\mathrm{net}$ fluctuates strictly within the statistical noise level $\pm 1/\sqrt{N}$ (shaded region), confirming the absence of a preferred direction.
The peak time of the TEs ($T_{Y \to X}$ and $T_{X \to Y}$) is slightly shorter than $\tau_\mathrm{th}$.
This shift could be attributed to the repulsive interaction $U_\mathrm{int}$, which accelerates the particles and consequently shortens the effective interaction time scale.
The situation is different from a two-skyrmion system confined within a square box~\cite{tani2026}, in which the TE peak positions coincide with $\tau_\mathrm{th}$.
In the two-skyrmion case, because of the gyrotropic force, the repulsive interaction results primarily in tangential circular motion rather than radial motion.

As the mass asymmetry is introduced and increased [Fig.~\ref{fig:te-components}(b) and (c)], the flow from the heavier particle $Y$ to the lighter particle $X$ ($T_{Y \to X}$) becomes larger than the reverse flow ($T_{X \to Y}$).
Consequently, $T_\mathrm{net}$ clearly emerges from the noise floor, forming a distinct positive peak. 
This explicitly visualizes the growth of the directed information transfer driven by the mass difference.
We can conclude that, from the heavier particle $Y$ which retains more information [higher AIS, Fig.~\ref{fig:AIS}(b)], net information flows to the lighter particle $X$.

In each panel of Fig.~\ref{fig:te-components}, the peak positions of all the three curves are almost the same.
As $\mu$ increases, the peaks are shifted to larger \rev{$\tau$}.
\rev{This behavior can be interpreted as a result of the increased absolute mass of particle $Y$, which slows down the system's dynamics and thereby delays the propagation of physical influence from the heavier particle to the lighter one.}

\begin{figure}
\begin{center}
   \includegraphics [width=0.85\linewidth]{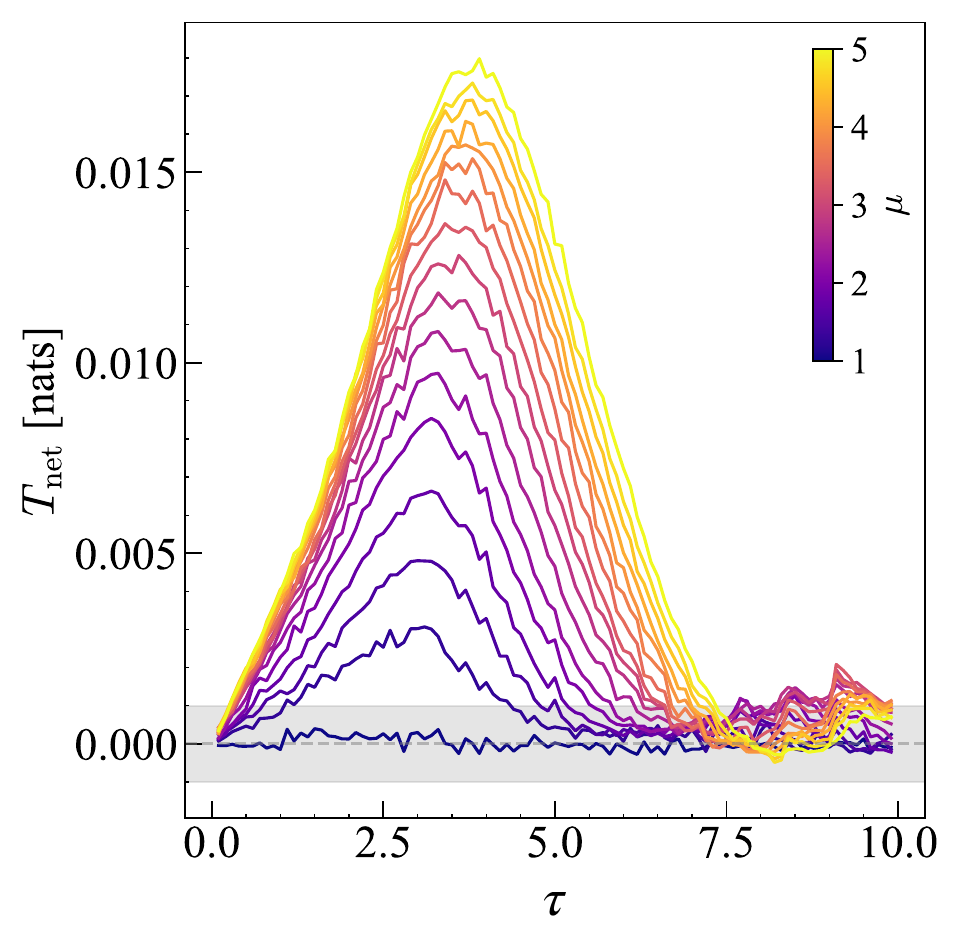}
   \caption{
   Net transfer entropy $T_\mathrm{net}$ for various mass ratios $\mu$, as a function of the time lag $\tau$.
   The shaded region represents the statistical noise $\pm 1/\sqrt{N}$.
            }
\label{fig:te-dynamics}
\end{center}
\end{figure}
To systematically capture this growth across a continuous range of mass asymmetry, Fig.~\ref{fig:te-dynamics} shows the net TE, $T_\mathrm{net}$, for various mass ratios $\mu$.
As $\mu$ increases, the peak position continuously shifts to larger \rev{$\tau$}.
Simultaneously, the peak height systematically increases, indicating that a heavier particle acts as a more robust source of information.
Comparing the mutual information (Fig.~\ref{fig:MI}) and the
transfer entropy (Fig.~\ref{fig:te-dynamics}) for various mass ratios $\mu$, we see that the decay of mutual information is faster than that of the TE peak.
This is because the random motion of either particle decreases the mutual information, as mentioned in Sec.~\ref{subsec:mi-ais}.

To explicitly understand what drives this robust peak in the net TE, we examine the mathematical connection between the memory difference (AIS) and the resulting directional information flow (net TE). 
From the definition of AIS [Eq.~\eqref{eq:ais-definition}], we have
\rev{
\begin{equation}
\begin{split}
    A_X(\tau) &= H(X_t) -H(X_t \mid X_{t-\tau}), \\
    A_Y(\tau) &= H(Y_t) -H(Y_t \mid Y_{t-\tau}).
\end{split}
\end{equation}
}
By using the definition of TE [Eq.~\eqref{eq:TE}], we can rewrite the Shannon entropies as
\rev{
\begin{equation}
\label{eq:shannon-decomposition}
\begin{split}
    H(X_t) &= A_X (\tau) + T_{Y\to X}(\tau) + H(X_t \mid X_{t-\tau}, Y_{t-\tau}),
    \\
    H(Y_t) &= A_Y (\tau) + T_{X\to Y}(\tau) + H(Y_t \mid X_{t-\tau}, Y_{t-\tau}).
\end{split}
\end{equation}
}
Assuming that the steady state is realized for sufficiently large $t$ (see Fig.~\ref{fig:shannon}), we write $H(X_t) = H(Y_t)$.
Therefore, we obtain
\rev{
\begin{equation}
\label{eq:T-net-ais}
    T_\mathrm{net}(\tau) = \delta A + \delta H_\mathrm{c},
\end{equation}
}
where we define
\rev{
\begin{equation}
\begin{split}
    \delta A &\equiv A_Y(\tau) - A_X(\tau), \\
    \delta H_\mathrm{c} &\equiv H(Y_t \mid X_{t-\tau}, Y_{t-\tau}) - H(X_t \mid X_{t-\tau}, Y_{t-\tau}).
\end{split}
\end{equation}
}
\rev{It should be noted that $H(X_t)$ and $H(Y_t)$ here represent the Shannon entropies associated with the positional degrees of freedom. 
Because the two particles are confined within the symmetric one-dimensional potential and coupled to the same thermal bath, their steady-state spatial probability distributions are symmetric to each other, yielding $H(X_t) = H(Y_t)$. 
In contrast, the model also possesses a measurable velocity degree of freedom.
Since the variance of the steady-state velocity distribution is given by $k_B T / m$, the velocity entropy explicitly depends on the mass and differs between the two particles. 
While the velocity degree of freedom implies an additional layer of information flow, its quantitative evaluation requires a significantly larger phase space and is beyond the scope of the present study. 
Here, we focus on the positional information to elucidate the fundamental spatial causality.}

\begin{figure}
\begin{center}
   \includegraphics [width=0.9\linewidth]{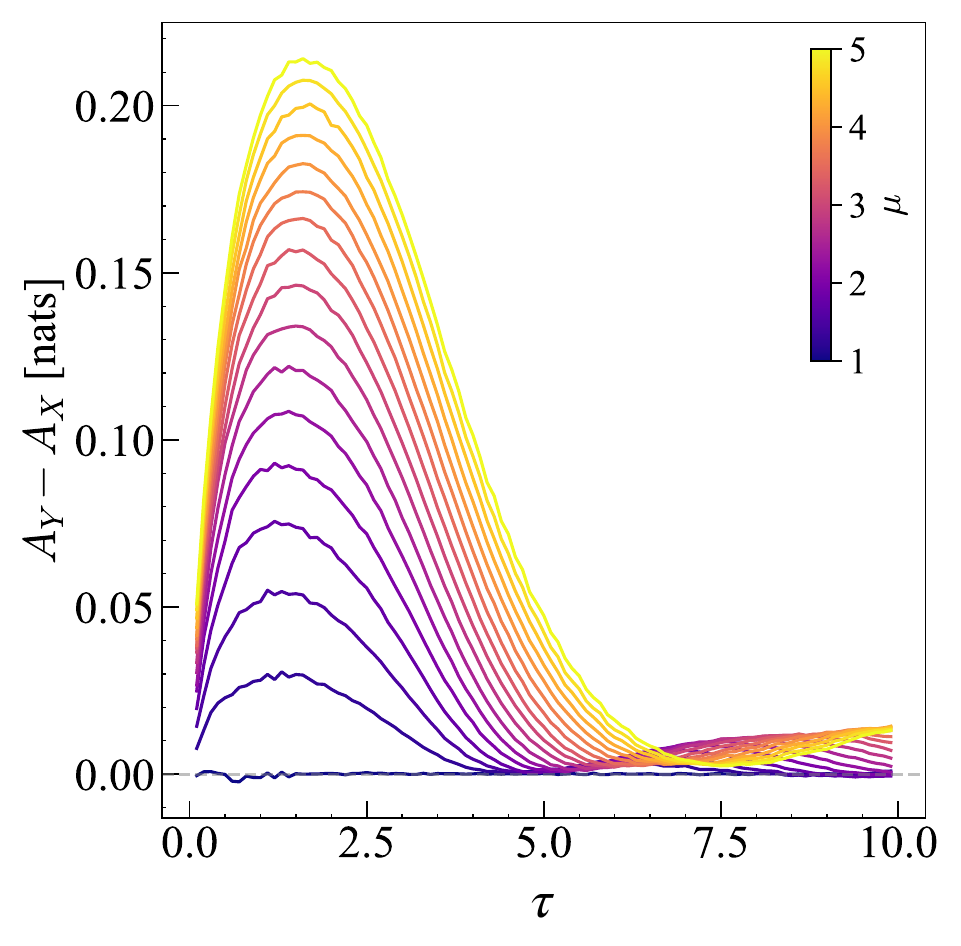}
   \caption{
   Difference of active information storage $\delta A = A_Y-A_X$ as a function of the time lag \rev{$\tau$}.
   The mass ratio $\mu$ varies from 1 to 5.
            }
\label{fig:ais-difference}
\end{center}
\end{figure}

Equation~\eqref{eq:T-net-ais} clearly states that the net TE consists of two components: the difference in AIS and in conditional Shannon entropy.
The first component \rev{$\delta A = A_Y(\tau) - A_X(\tau)$} is illustrated in Fig.~\ref{fig:ais-difference}, for several mass ratios $\mu$.
The shapes of these curves closely resemble the structures of the net TE, \rev{$T_\mathrm{net}(\tau)$} (Fig.~\ref{fig:te-dynamics}), directly supporting our claim that the difference in memory capacity (AIS) inherently drives the directional information propagation.

However, a comparison between Fig.~\ref{fig:te-dynamics} and Fig.~\ref{fig:ais-difference} reveals a significant difference in magnitude: the AIS difference is approximately an order of magnitude larger than the net TE.
This difference is precisely compensated by the second component in Eq.~\eqref{eq:T-net-ais}, $\delta H_\mathrm{c}$.
Given the past states of both particles \rev{$(X_{t-\tau}, Y_{t-\tau})$}, the heavier particle $Y$, due to its larger inertia, tends to exhibit a more predictable trajectory compared to the lighter particle $X$.
Consequently, the conditional uncertainty of $Y$ is strictly smaller than that of $X$, resulting in $\delta H_\mathrm{c} < 0$.
Thus, this negative $\delta H_\mathrm{c}$ term strongly decreases the raw memory difference $\delta A$, leaving the smaller yet distinct positive net information flow $T_\mathrm{net}$ from the heavier to the lighter particle.

\begin{figure}
\begin{center}
   \includegraphics [width=0.85\linewidth]{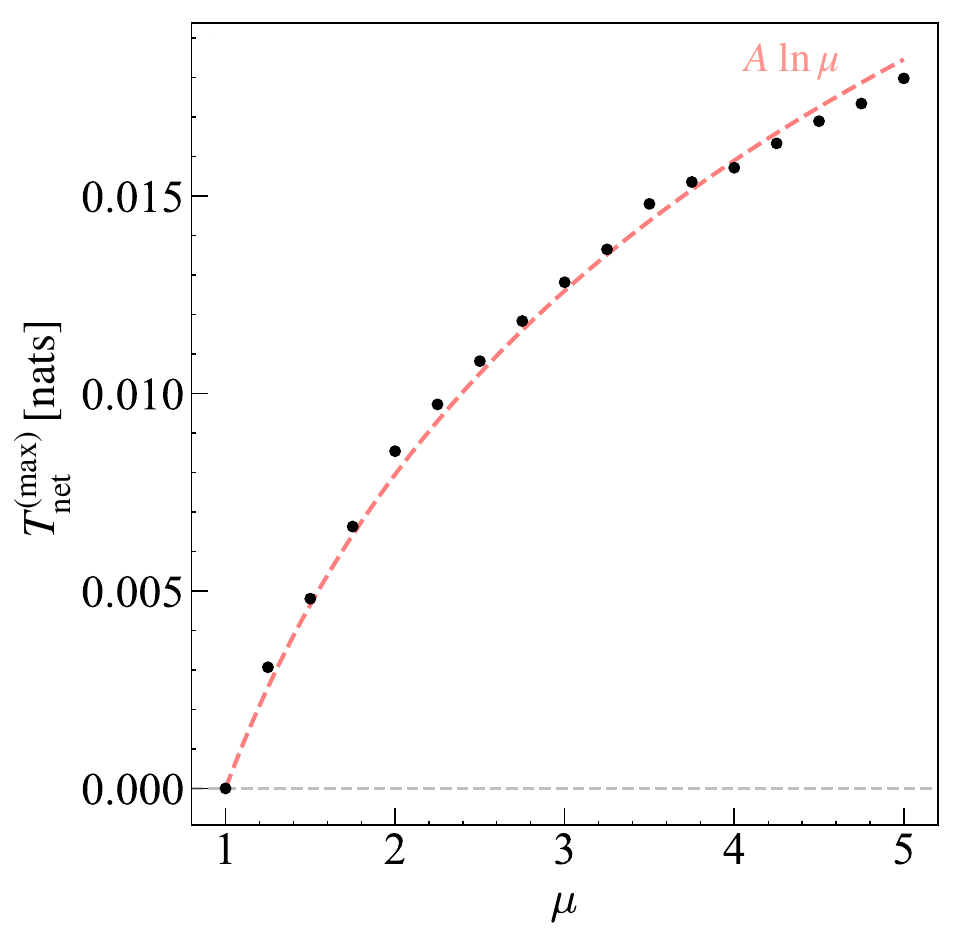}
   \caption{
   Maximum value of the net transfer entropy, $T_\mathrm{net}^{\mathrm{(max)}}$, plotted against the mass ratio $\mu$.
   The dashed curve is a logarithmic fit $f(\mu)=A\ln{\mu}.$
            }
\label{fig:te-scaling}
\end{center}
\end{figure}
To further characterize the impact of mass asymmetry on the directional transfer of information, we extract the maximum value of the net TE, 
\rev{
\begin{equation}
    T_\mathrm{net}^{\mathrm{(max)}} = \max_{\tau} \, [T_\mathrm{net} (\tau)],
\end{equation}
}
and plot it against the mass ratio $\mu$ in Fig.~\ref{fig:te-scaling}.
The plot clearly reveals that the information flow grows with a decreasing rate as the mass asymmetry increases.
To minimally capture this scaling behavior, we apply a \rev{simple empirical} logarithmic fit, $f(\mu) = A \ln \mu$.
As shown in Fig.~\ref{fig:te-scaling}, this simple model well describes the overall trend within the currently investigated range ($1 \le \mu \le 5$) with the fitting parameter $A=0.0115$.
Note that the slight scatter of the data points around the fitting curve (e.g., at $\mu=3.5$) originates from the inherent statistical fluctuations in the TE.
As observed in Fig.~\ref{fig:te-dynamics}, the net TE, $T_\mathrm{net}(\tau)$, exhibits minor jaggedness near its peak due to the estimation of joint probabilities from the finite ensemble size ($N=10^6$).
Consequently, extracting the absolute maximum value occasionally captures these local noise spikes rather than the ideal peak value, leading to the observed minor deviations.
Nevertheless, these statistical variations are sufficiently small and do not obscure the overall \rev{empirical trend} of $T_\mathrm{net}^{\mathrm{(max)}}$.

\rev{While other saturating functions could also potentially fit this trend, this empirical logarithmic scaling serves as a simple description} of the saturation process towards a deterministic limit.
As $\mu \to \infty$, the heavier particle becomes increasingly deterministic, causing its capacity to transfer causal information to eventually hit an upper bound.
Thus, the logarithmic function naturally captures this gradual saturation of information flow driven by the extreme mass asymmetry.

In the preceding analyses, we employed a spatial discretization of $N_\mathrm{cell} = 4$ cells for two primary reasons.
First, as $N_\mathrm{cell}$ increases, the computational and data requirements for entropic quantities dramatically escalate. 
Specifically, evaluating the TE [Eq.~\eqref{eq:TE}] requires estimating the joint probability distribution over $(N_\mathrm{cell})^3$ states. 
Consequently, the ensemble size $N$ required to secure statistical accuracy and suppress finite-sampling biases grows cubicly. 
While simulating a larger $N_\mathrm{cell}$ is theoretically possible, obtaining such massive datasets is often unrealistic in actual experiments. 
We deliberately chose $N_\mathrm{cell}=4$ to align with practically feasible experimental setups~\cite{suzuki2025informationdynamicsnaturalcomputing}.
Second, if we were to increase $N_\mathrm{cell}$, the primary effect would be a quantitative shift of the TE peak to a smaller \rev{$\tau$}. 
For example, $N_\mathrm{cell} = 8$ would yield a peak position slightly shorter than $\tau_\mathrm{th}/2$, simply because the spatial distance between adjacent cells is halved ($\xi/2$), thus requiring less time for a particle to traverse. 
Therefore, although the peak position shifts quantitatively with the spatial resolution, our qualitative conclusion regarding the underlying mechanism of directional information flow remains robust and unaffected by the choice of $N_\mathrm{cell}$.

\rev{
Furthermore, it is theoretically important to consider the continuous limit ($N_\mathrm{cell} \to \infty$). 
Because the TE is defined as a difference of conditional entropies, the singular terms $\ln(\Delta x)$ that arise from the continuous limit of discrete spatial states exactly cancel out. 
Thus, a physically relevant continuous limit (the differential transfer entropy evaluated via probability density functions) is well-defined and finite. 
Regarding the magnitude of the net TE, spatial coarse-graining into finite cells inevitably obscures microscopic dynamics; causal influences that cause a particle to move but not cross a cell boundary remain unrecorded. 
Consequently, the magnitude of the net TE evaluated with a finite $N_\mathrm{cell}$ serves as a lower bound. 
As $N_\mathrm{cell} \to \infty$, the finer resolution captures these hidden microscopic influences, causing the peak magnitude to asymptotically increase toward the continuous TE value, while the peak position $\tau$ simultaneously shifts toward zero.
}

\rev{
It is instructive to consider the relationship between the information-theoretic measures investigated here and general thermodynamic quantities. 
In the framework of information thermodynamics, the TE bounds the entropy production~\cite{ito2015informationflowentropyproduction,parrondo2015thermodynamics}.  
This connection is particularly relevant in the transient regime [e.g., Fig.~2(b)], where the information transfer dictates the minimum thermodynamic cost of the transient heat flow during relaxation. 
Regarding the AIS, one plausible implication arises from the thermodynamics of prediction~\cite{Still2012}, which suggests that a system's ability to retain predictive information is related to the lower bound of dissipated work. 
In our system, the heavier particle's high AIS reflects its robustness against thermal fluctuations. 
This suggests that AIS could be physically interpreted as the capacity of a system to maintain temporal coherence, potentially dictating the thermodynamic efficiency of memory retention and dissipation. 
Examining these relationships between information-theoretic measures and thermodynamic quantities may be an interesting direction for future studies.
}

\section{Conclusion}
\label{sec:conclusion}
In conclusion, we have theoretically investigated the directional information flow between two interacting Brownian particles confined in a one-dimensional potential. 
By simulating the coupled Langevin equations and evaluating TE, we demonstrated that a mass asymmetry inherently breaks the symmetry of information exchange. 
Specifically, we revealed that a net information flow emerges, directed consistently from the heavier particle to the lighter one. 
We found that the TE value increases when the mass ratio increases.
This behavior can be understood as follows: the directional information transfer is physically rooted in the difference in inertia. 
The heavier particle, being less susceptible to thermal fluctuations, retains the memory of its past trajectory for a longer duration, as evidenced by its higher AIS.
Consequently, it acts as a stable source of information that continuously influences the stochastic motion of the lighter particle. 
We analytically pointed out a mathematical connection between the directional information flow (net TE) and the memory (AIS).
Specifically, our decomposition of the net TE reveals that this directional flow is governed by a competition between the difference in raw memory capacity ($\delta A$) and the difference in predictability of the particle trajectories ($\delta H_\mathrm{c}$).

The TE peak position is slightly shifted from the thermal characteristic time $\tau_\mathrm{th}$, unlike the case of a two-skyrmion system confined within a square box~\cite{tani2026}.
In the two-skyrmion system, a skyrmion--skyrmion interaction results in tangential circular motion; in the case of the one-dimensional Brownian particles, a particle--particle interaction accelerates the particles and gives a TE peak position shorter than $\tau_\mathrm{th}$.

Furthermore, as demonstrated in our analysis of mutual information, conventional symmetric measures only capture the shared mechanical periodicity of the system and fail to detect this causal directionality. 
Thus, evaluating TE is essential for isolating the fundamental mechanical effects on information propagation.

\rev{We established that the maximum net TE exhibits a sublinear growth with respect to the mass ratio, which can be empirically approximated as $T_{\mathrm{net}}^{\mathrm{(max)}} \propto \ln \mu$ across the investigated parameter range.
This empirical approximation indicates that the logarithmic function serves as a convenient simple description to capture the saturating trend of the net TE $T_\mathrm{net}^\mathrm{(max)}$ on the mass ratio $\mu$; the net TE is physically expected to saturate towards a certain finite value for $\mu\to\infty$.}

Our results reveal a clear physical interpretation of TE.
More importantly, these findings provide a fundamental understanding of information flow in general physical systems, including nanoparticles and molecules that exhibit Brownian motion.
These insights into directional information flow pave the way to analyze more complex systems such as interacting skyrmion bits driven by topological gyrotropic motion, which hold significant promise for future applications in stochastic computing and nanoscale information processing.

\rev{
While the present study focuses on massive Brownian particles where directionality is dictated by mass asymmetry, it is relevant to consider the overdamped limit. 
In the absence of mass, an asymmetry in particle size leads to different viscous damping coefficients $\gamma$. 
Because a larger particle experiences higher friction, it exhibits a slower relaxation timescale and higher AIS compared to a smaller particle. 
Consequently, even in the overdamped regime, we expect that size asymmetry could induce a net information transfer from the larger (slower) particle to the smaller (faster) particle. Exploring this directional information flow in interacting overdamped systems, such as colloidal suspensions, could be an intriguing direction for future experimental and theoretical research.
}

{
Beyond the specific model of interacting Brownian particles, our findings offer a fundamental perspective on the interpretation of TE in general non-Markovian systems. 
In the analysis of complex dynamics, a directional information flow is frequently attributed to structural asymmetries, such as unidirectional couplings~\cite{Schreiber2000,Halavackova2007,Biswajit2025} or feedback mechanisms~\cite{ito2013information,horowitz2014}. 
However, our results demonstrate that causal directionality can inherently emerge even in systems with perfectly symmetric interactions, driven purely by the asymmetry in memory capacities (AIS). 
This physically originates from the difference in inertia; purely mechanical differences can manifest as robust causality.
This implies that when inferring causal relationships from time-series data, it is crucial to account for the disparate physical timescales or ``inertia" of individual components. 
Our study thus provides a vital physical baseline for TE analyses of general non-Markovian complex systems.
}

\begin{acknowledgements}
This work was supported by Japan JSPS KAKENHI Grant No.~JP23KJ1497. 
I thank Eiiti Tamura, Yoshishige Suzuki, and Soma Miki for fruitful discussions.
\end{acknowledgements}

\appendix
\section{Correlation function}
\label{app:corr}
In this appendix, we numerically show that the correlation function cannot capture causal information flow.
We define the fluctuations of the discrete spatial states from their ensemble averages as
\begin{equation}
\begin{split}
    & \delta a_t^{(i)} = a_t^{(i)} - \langle a_t \rangle,
    \\
    & \delta b_t^{(i)} = b_t^{(i)} - \langle b_t \rangle.
\end{split}
\end{equation}
The correlation function is then given by
\rev{
\begin{equation}
    C(\tau) = \frac{1}{N} \sum_{i=1}^{N} \delta a_t^{(i)} \delta b_{t-\tau}^{(i)},
\end{equation}
}
which vanishes for sufficiently large time delay \rev{$\tau$}.

\begin{figure}
\begin{center}
   \includegraphics [width=0.85\linewidth]{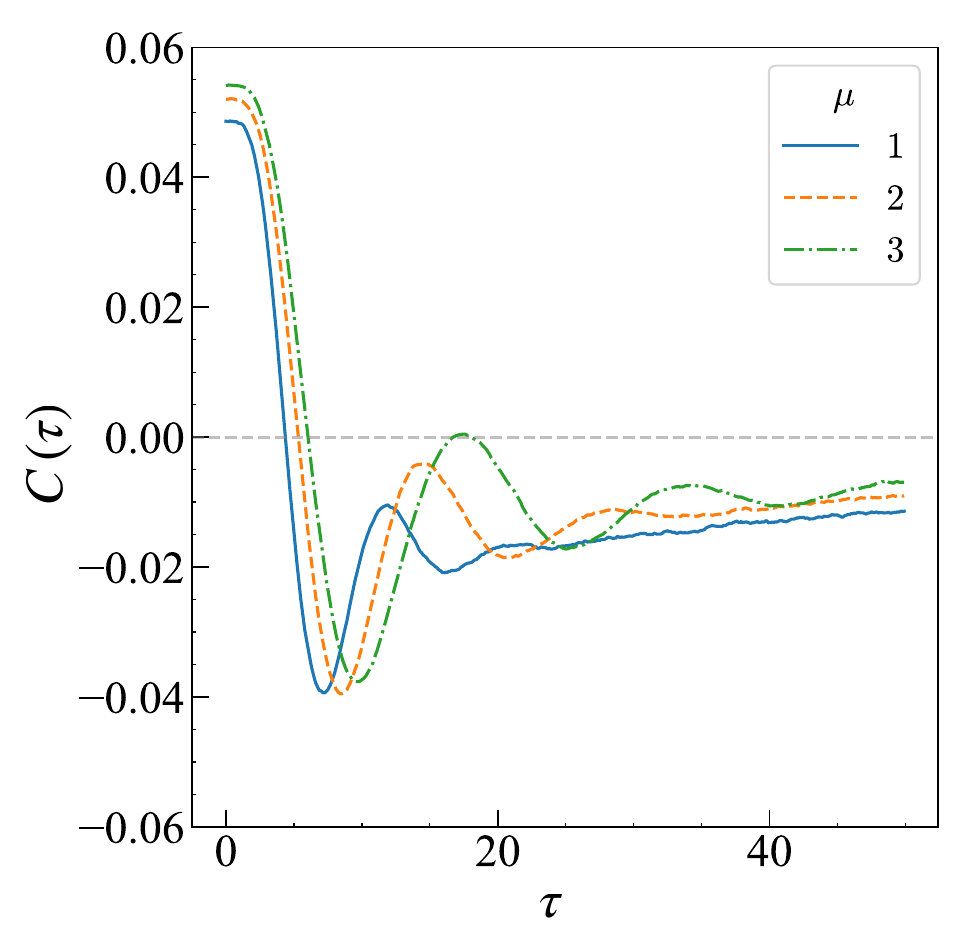}
   \caption{
   Correlation functions \rev{$C(\tau)$} for mass ratios $\mu =1,2,$ and $3$.
            }
\label{fig:correlation}
\end{center}
\end{figure}
Figure~\ref{fig:correlation} illustrates the correlation functions \rev{$C(\tau)$} for both symmetric ($\mu=1$) and asymmetric ($\mu=2, 3$) cases with $t=t_\mathrm{f}$.
The curves exhibit damped oscillations characterized by alternating positive peaks and negative valleys.
These oscillatory structures merely reflect the periodic bouncing of the particles within the confinement potentials [see Fig.~\ref{fig:shannon} in the main text]. 
As $\mu$ increases, the oscillation period lengthens simply due to the larger inertia of the heavier particle.
Crucially, these peaks appear even in the perfectly symmetric case ($\mu=1$) where no net directional information flow can physically exist. 
Furthermore, increasing the mass asymmetry ($\mu > 1$) does not generate any distinct feature representing causal directionality.
These results clearly demonstrate that the conventional correlation function only captures the shared mechanical periodicity and is fundamentally incapable of isolating the true causal asymmetry and directional information transfer driven by the mass difference.

\section{Derivation of nondimensionalized equations of motion}
\label{app:derivation}

\rev{
In the main text, we nondimensionalize the Langevin equations [Eq.~\eqref{eq:langevin}] to simplify the form and concentrate on the fundamental physics.
Here, we give a detailed description of this procedure.
We choose $m_X$, $U_0$, and $\xi$ as the basic units of mass, energy, and length, respectively.
Consequently, the characteristic unit of time is defined as $t_0 \equiv \xi \sqrt{m_X/U_0}$.
Equation \eqref{eq:langevin} is then simplified as
\begin{equation}
    \begin{split}
        \dv[2]{\hat{x}_X}{\hat{t}} = -\hat{\gamma} \dv{\hat{x}_X}{\hat{t}} &+ e^{-|\hat{x}_1-\hat{x}_2|}\sgn{(\hat{x}_X - \hat{x}_Y)} \\
        &+ \frac{\hat{U}_{\mathrm{w}0}}{\hat{\xi}_\mathrm{w}}[e^{-\hat{x}_X/\hat{\xi}_\mathrm{w}} - e^{-(\hat{L}-\hat{x}_X)/\hat{\xi}_\mathrm{w}}] + \hat{R}_X,
        \\
        \mu \dv[2]{\hat{x}_Y}{\hat{t}} = -\hat{\gamma} \dv{\hat{x}_Y}{\hat{t}} &- e^{-|\hat{x}_X-\hat{x}_Y|}\sgn{(\hat{x}_X - \hat{x}_Y)} \\
        &+ \frac{\hat{U}_{\mathrm{w}0}}{\hat{\xi}_\mathrm{w}}[e^{-\hat{x}_Y/\hat{\xi}_\mathrm{w}} - e^{-(\hat{L}-\hat{x}_Y)/\hat{\xi}_\mathrm{w}}] + \hat{R}_Y,
    \end{split}
    \label{eq:langevin-dimless}
\end{equation}
where $\hat{x}_X \equiv x_X / \xi, \hat{x}_Y \equiv x_Y / \xi$ are the dimensionless position, and $\hat{t} \equiv t/t_0$ is the dimensionless time.
We define dimensionless parameters $\hat{\gamma} \equiv \gamma (\xi / \sqrt{m_X U_0})$, $\hat{U}_{\mathrm{w}0} \equiv U_{\mathrm{w}0}/U_0$, $\hat{\xi}_\mathrm{w} \equiv \xi_\mathrm{w}/\xi$, and $\hat{L} = L/\xi$.
To introduce mass asymmetry, we set $m_X\neq m_Y$, which yields a mass ratio $\mu \equiv m_Y/m_X \neq 1$.
The dimensionless random force $\hat{R}_j$ satisfies the fluctuation--dissipation theorem and is discretized in our numerical simulations as
\begin{equation}
    \hat{R}_X,\hat{R}_Y \sim \sqrt{\frac{2\hat{\gamma} \hat{T}}{\delta \hat{t}}} \mathcal{N}(0,1),
\end{equation}
where $\mathcal{N}(0,1)$ the standard normal distribution.
Here, $\hat{T} \equiv k_\mathrm{B}T /U_0$ is the dimensionless temperature, and $\delta\hat{t}$ represents the discrete dimensionless time step.
}

\bibliography{reference}
\end{document}